\documentclass[11pt,a4paper]{article}
\usepackage[utf8]{inputenc}
\usepackage{amsmath}
\usepackage{amsfonts}
\usepackage{amssymb}
\usepackage{graphicx}
\usepackage{bm}
\usepackage{booktabs}       
\usepackage{multirow}
\usepackage{epstopdf}
\usepackage{hyperref}       
\usepackage{url}            
\usepackage[top=2.52cm, bottom=2.52cm, left=2.52cm, right=2.52cm]{geometry}
\usepackage{xcolor} 
\usepackage{setspace}
\begin{document}
\title{\textbf{Perfect absorption of molecular vibration enabled by critical coupling in molecular metamaterial}}



\author{Govind Dayal\footnote{gdayal@physics.du.ac.in} $^{, 1}$, Dheeraj Pratap\footnote{dpratap@iitd.ac.in} $^{, 2}$\\ \\
$^{1}$Department of Physics and Astrophysics, University of Delhi, Delhi 110007, India\\
	$^{2}$Optics and Photonic Centre, Indian Institute of Technology Delhi, Delhi 110016, India\\
	}

\date{}

\maketitle

\begin{abstract}
The absorption and emission spectrum arising from the vibrational motion of a molecule is mostly in the infrared region. These fingerprint absorptions of polar bonds enable us to acquire bond-specific chemical information from specimens. However, the mode mismatch between the atomic-scale dimensions of the chemical bonds and the resonance wavelength limits the direct detection of tiny amounts of samples such as self-assembled monolayers or biological membranes. To overcome this limitation, surface-enhanced infrared absorption spectroscopy (SEIRA) has been proposed to enhance infrared absorption directly via local field enhancement. Here, we report on the perfect absorption of molecular vibration enabled by critical coupling in the metamaterials. Our molecular metamaterial design consists of a thin polymer layer sandwiched between a structured metal layer on top and a continuous metal layer at the bottom that supports the gap plasmon mode. The measured and simulated infrared spectra of the molecular metamaterial show broad and narrow absorption bands corresponding to the metamaterial and molecular vibration modes. We show that by tuning the structure's molecular film thickness and periodicity, vibrational absorption can be enhanced to near unity. We also show that for a particular periodicity of the array, metamaterial resonance can be completely suppressed, and only molecular vibrational absorption is excited, giving rise to an extremely narrow absorption band.

\vspace{3mm}
\end{abstract}


\section{Introduction}
Metamaterials, which are composite mediums of artificially designed subwavelength resonant particles (sometimes called meta-atoms or nanoantennas), have revolutionized the study of light-matter interaction by providing a fully controllable platform for light manipulation\cite{SAR2008, Pendry2006, Cui2024}. One of the most sought-after applications of metamaterials is in light confinement and light harvesting, where high absorption of electromagnetic waves is required, e.g., molecular sensing, photocurrent generation, and photodetection\cite{Watts2012, Charlene2019}. In metamaterials, near-perfect absorption is achieved by tailoring the electric and magnetic resonances of its inclusions in such a way that the effective index of the homogenized medium matches with the free space. The most popular and fabrication-friendly design of highly absorbing metamaterials is a tri-layer metal-insulator-metal design with the top metal layer structured to act as an electric dipole resonator that determines the electric resonance of the medium\cite{Landy2008}. The magnetic resonance of the medium is excited with the help of image charges of an electric dipole formed in the bottom metal layer, which is also known as the ground plane. The resultant circulating currents yield a magnetic dipole character response that is associated with a reduction in radiation damping. 

At a critical spacing thickness $d$, the radiation from these two induced dipoles is perfectly out of phase \cite{Howard1998}, leading to the cancelation of reflection. This condition is called the critical coupling condition and can be understood with the help of cavity losses~\cite{Yoon2010,Emeric2019}. The reflection coefficient of the tri-layer metamaterial can be expressed using the coupled mode theory as
\begin{equation}
\centering
r(\omega) = \frac{({\gamma_{\mathrm{rad}}-\gamma_{\mathrm{abs}}})-i({\omega-\omega_{0}})}{({\gamma_{\mathrm{rad}}+\gamma_{\mathrm{abs}}})+i({\omega-\omega_{0}})}
\end{equation}
where $\omega_{0}$ is the resonant frequency of the eigenmode and $\gamma_{\mathrm{rad}}$, $\gamma_{\mathrm{abs}}$ represent the radiative and the non-radiative losses, respectively. In tri-layer metamaterials, the thickness of the spacer layer provides an easy method to precisely tune the radiative loss of the system and $\gamma_{\mathrm{rad}}$ can be engineered to match $\gamma_{\mathrm{abs}}$ of the composite medium such that $r(\omega)$ goes to zero. 
Moreover, the ground plane, which is an optically thick metal film, prevents any transmission of radiation, and thus most of the electromagnetic energy is confined in the insulating medium. The fundamental mode of the metamaterial is characterized by confined electromagnetic fields within the gap, with an enhanced electric field near the edges and a maximum magnetic field at the centre of the cavity. This deep sub-wavelength confinement of energy gives rise to drastically enhanced electric and magnetic fields that can be exploited to study the strong light-matter interactions across the spectrum.

In the mid-infrared range, field enhancement can be exploited to directly enhance the infrared absorption of molecular vibrations that depend on the coupling between enhanced local electric fields and the fingerprint vibrations of the molecules of interest\cite{Neubrech2017, Kozuch2023}. This is done by tuning the metamaterial resonances to overlap with the absorption bands of targeted molecules. In general, metamaterial resonances are quite broad compared to molecular absorption bands because of strong radiative and nonradiative losses. Thus, interference of broad-band and narrow-band resonances results in an asymmetric line-shaped resonance, known as the Fano resonance. In most cases, the molecules to be detected are placed on top of the metamaterial surface, where they interact with the localized field near the edges of the electric dipole resonators. However, as discussed before, the maximum field enhancement is inside the spacer, so one needs to place the molecules in the gap, which is not possible because the gap is occupied by the insulating spacer layer. However, the hotspots in the gap can be accessed by modifying the geometry of the trilayer structure using the pedestal geometry\cite{Cetin2014, Hwang2018}. In this case, the molecules can interact with the electric field with maximum enhancement, and the sensitivity of detection can be increased further. It is important to note that an array of electric dipoles maximally absorbs 50 $\%$ of the impinging light\cite{Alaee2017}. Therefore, to the best of our knowledge, perfect absorption of molecular vibrations has not been achieved so far. In this work, we numerically demonstrate perfect absorption of the C=O stretch mode of PMMA molecules by adjusting the periodicity of the array and the thickness of the polymer film to obtain the critical coupling condition. 

\begin{figure}[t]
\centering
\includegraphics[width=14cm]{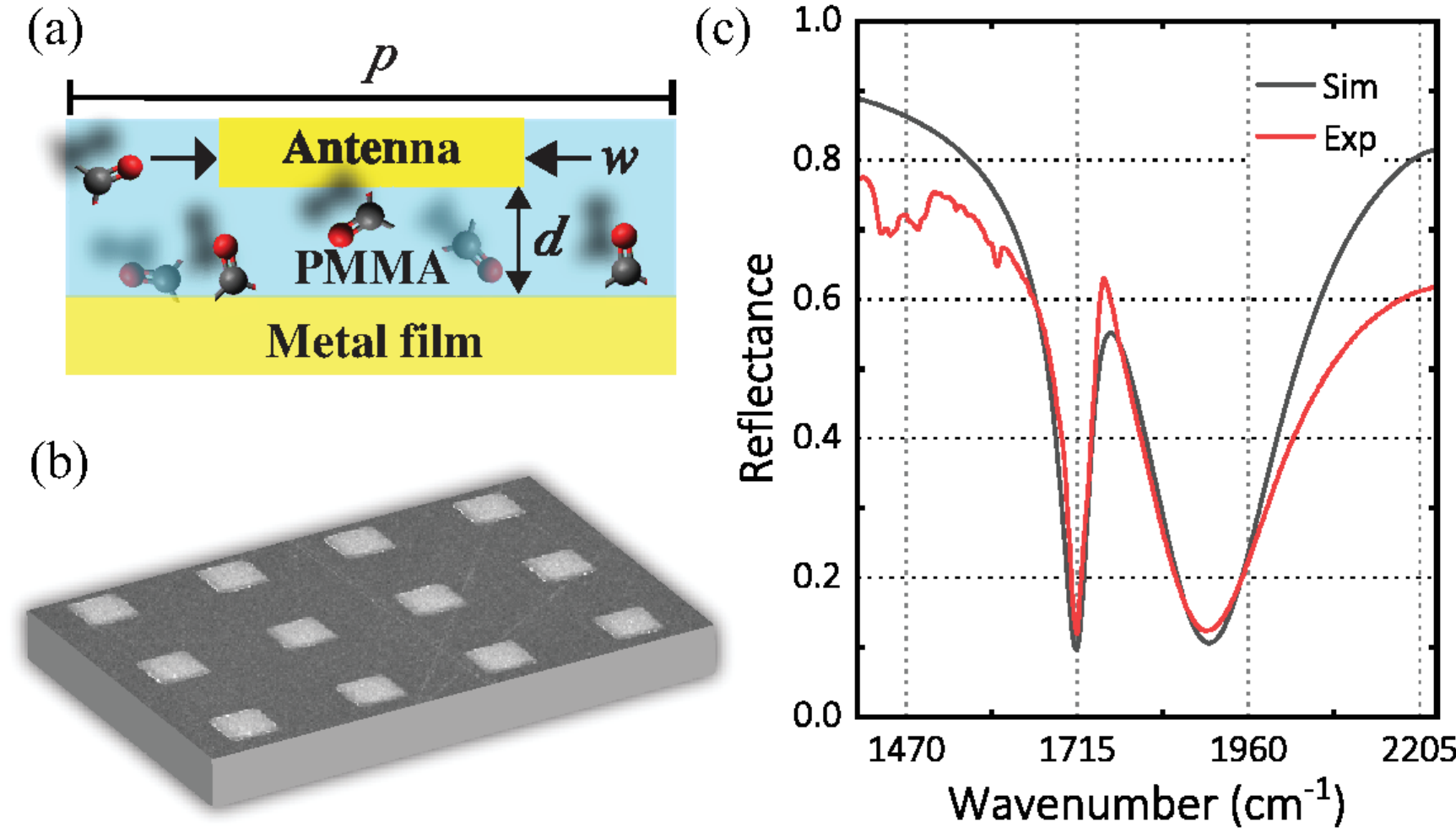}
\caption{(a) Schematic cross-sectional view of the metamaterial perfect absorber made of antenna array on top of a metallic ground plate separated by a PMMA layer. Parameters of molecular metamaterial include patch width $w$, PMMA thickness $d$, and periodicity of the square array $p$. (b) Scanning electron microscopic image of the fabricated sample. (c) Experimental and simulated reflection spectra of molecular metamaterial with periodicity $p$ = 2.8 ${\mu}$m, PMMA film thickness $d$ = 160 nm and patch width $w$ = 1.2 ${\mu}$m.}
\end{figure}


The proposed molecular metamaterial consists of a PMMA layer sandwiched between a metal ground plane and a patterned array of metallic nanostructures, as shown in Fig. 1 (a). The size of the square metallic patches, which define the resonant wavelength, is precisely controlled by electron-beam lithography. Experimentally, a 1 mm$^2$ square array of nanopatches was patterned onto a CaF$_2$ substrate by electron beam lithography, metal evaporation, and lift-off. We start the fabrication by spin-coating a layer of approximately 400 nm ZEP 520A to serve as a positive-tone electron-beam resist layer. A thin conductive polymer layer (E-Spacer 300Z) is further spin-coated to mitigate charging effects during electron-beam lithography. After the exposer, the resist was developed using a ZED-N50 developer for 60 s and rinsed in ZMD-B for 30 s. A 5-nm thick Cr layer and an 80-nm Au layer were deposited on the developed sample, followed by the lift-off to form a nanoantenna array. Au is used for both metal layers and its thickness is approximately 100 nm. The remaining resist is finally removed by rinsing the sample in acetone overnight. The SEM micrograph of the nanopatch array is shown in Fig. 1(b).

In Fig. 1 (c), we plot the experimental and simulated reflectance spectra of the proposed molecular metamaterial. Reflection spectra are measured for unpolarized light with 4 cm$^{-1}$ resolution using a FTIR spectrometer (VERTEX 70v, Bruker). The hybrid system shows two dips at 1900 cm$^{-1}$ and 1732 cm$^{-1}$ that are due to the excitation of the gap plasmon mode (refer to metamaterial resonance from hereafter) and the vibrational mode associated with the C=O stretch of the PMMA molecule (refer to molecular resonance from hereafter). The gap plasmon mode is characterized by the excitation of electric and magnetic dipole modes in the metal-insulator-metal cavity as shown in Figs.~2(a) and Fig.~2(b) \cite{Dayal2012}. The induced electric dipole moment $\mathrm{p_{x}}$ in the nanoantenna is expressed as 
\begin{equation}
\centering
\mathrm{\textit{p}_{x}}=\epsilon_{0}\mathrm{\alpha^{eff}_{ee}}\mathrm{\textit{E}^{inc}_{x}}
\end{equation}
where $\mathrm{\alpha^{eff}_{ee}}$ is the effective electric polarizability and $\epsilon_{0}$ is the free space permittivity. Similarly, the magnetic dipolar moment can be written as
\begin{equation}
\centering
\mathrm{\textit{m}_{y}}=\mathrm{\alpha^{eff}_{mm}}\mathrm{\textit{H}^{inc}_{y}}
\end{equation}
where $\mathrm{\alpha^{eff}_{mm}}$ is the effective magnetic polarizability. The expression for the magnetic polarizability of the metamaterial was analytically derived by Bowen et. al \cite{Bowen2014,Bowen2016}, and it is given by

\begin{equation}
\centering
\mathrm{\alpha^{eff}_{mm}}=\frac{8dc^{2}cos^2({sin(\theta})kw_{\mathrm{eff}}/2)}{\omega^{2}_{0}-\omega^{2}+i\omega^{2}/Q}
\end{equation}
where $w_{\mathrm{eff}}$ is the effective width of antenna, $d$ is gap thickness and $\theta$ is the angle of the incident magnetic field. $1/Q = 1/Q^{\mathrm{rad}}+1/Q^{\mathrm{abs}}$ is the $Q$ factor of the mode due to both radiative and Ohmic losses. The $Q^{\mathrm{rad}}$ is defined by
\begin{equation}
\centering
Q^{\mathrm{rad}} = \frac{kp^{2}cos(\theta)}{2d(1-r_{TM})A_{\mathrm{t}}(\theta)}
\end{equation}
where $A_{\mathrm{t}}(\theta)$ is $cos^{2}[sin(\theta)kw_{\mathrm{eff}}/2]$ and $r_\mathrm{TM}$ is the reflection coefficient for a transverse magnetic (TM) wave. It has been shown that maximum absorption occurs when 
$\mathrm{\alpha^{eff}_{mm}}$ = $\mathrm{\alpha^{eff}_{ee}}$. This condition is called the balanced Kerker condition for effective polarizabilities\cite{Alaee2017}. As we can see from Equation 5, the radiative loss of the composite medium depends strongly on the periodicity of the array and the spacer layer thickness. This suggests that both periodicity and spacer layer thickness can be adjusted to obtain the critical coupling condition to achieve perfect absorption of radiation.

\begin{figure}[t]
\centering
\includegraphics[width=13cm]{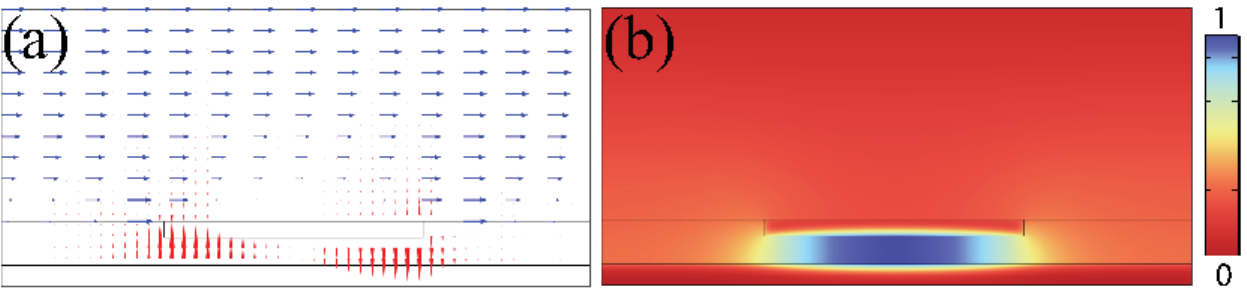}
\caption{Simulated electric and magnetic field in the molecular metamaterial at 1900 cm$^{-1}$. The blue arrows in (a) show the incident electric field, while the red arrows show the field within the spacer layer. The electric field within the gap region shows the formation of two out-of-phase induced dipoles. These two dipoles lead to a magnetic moment, indicated with the color plot in Fig. (b).}
\end{figure}

To understand the nature of metamaterial and molecular absorption modes and their dependence on different geometrical parameters, we performed comprehensive and parameterized electromagnetic simulations (COMSOL Multiphysics)\cite{COMSOL} based on the Finite-Element Method (FEM). Periodic boundary conditions in the in-plane dimensions and perfectly matched layers in the out-of-plane dimension were used throughout with excitation due to a normally incident plane wave. The metallic parts of our metamaterial resonator were modeled using a Drude model with a plasma frequency of $\omega_{\mathrm{p}}$=2$\pi\times$2175 THz and collision frequency of $\gamma$=2$\pi\times$6.5 THz. The PMMA permittivity describing the C=O stretch was modeled as a Lorentz oscillator $\varepsilon_\mathrm{m} = \varepsilon_\mathrm{b}+f\omega_\mathrm{m}^{2}/(\omega_\mathrm{m}^{2}-i\gamma_\mathrm{m}\omega-\omega^{2}),$ where the relative permittivity $\varepsilon_\mathrm{b}$ is 2.2, the oscillator strength $f$ is 0.01, the resonance frequency $\omega_\mathrm{m}$ is 1732 cm$^{-1}$, and $\gamma_\mathrm{m}$ is 20 cm$^{-1}$ \cite{Dayal2021}. We numerically calculated the S-parameter for all structures with normal incidence to obtain the reflectivity spectra. 

\begin{figure}[t]
\centering
\includegraphics[width=14cm]{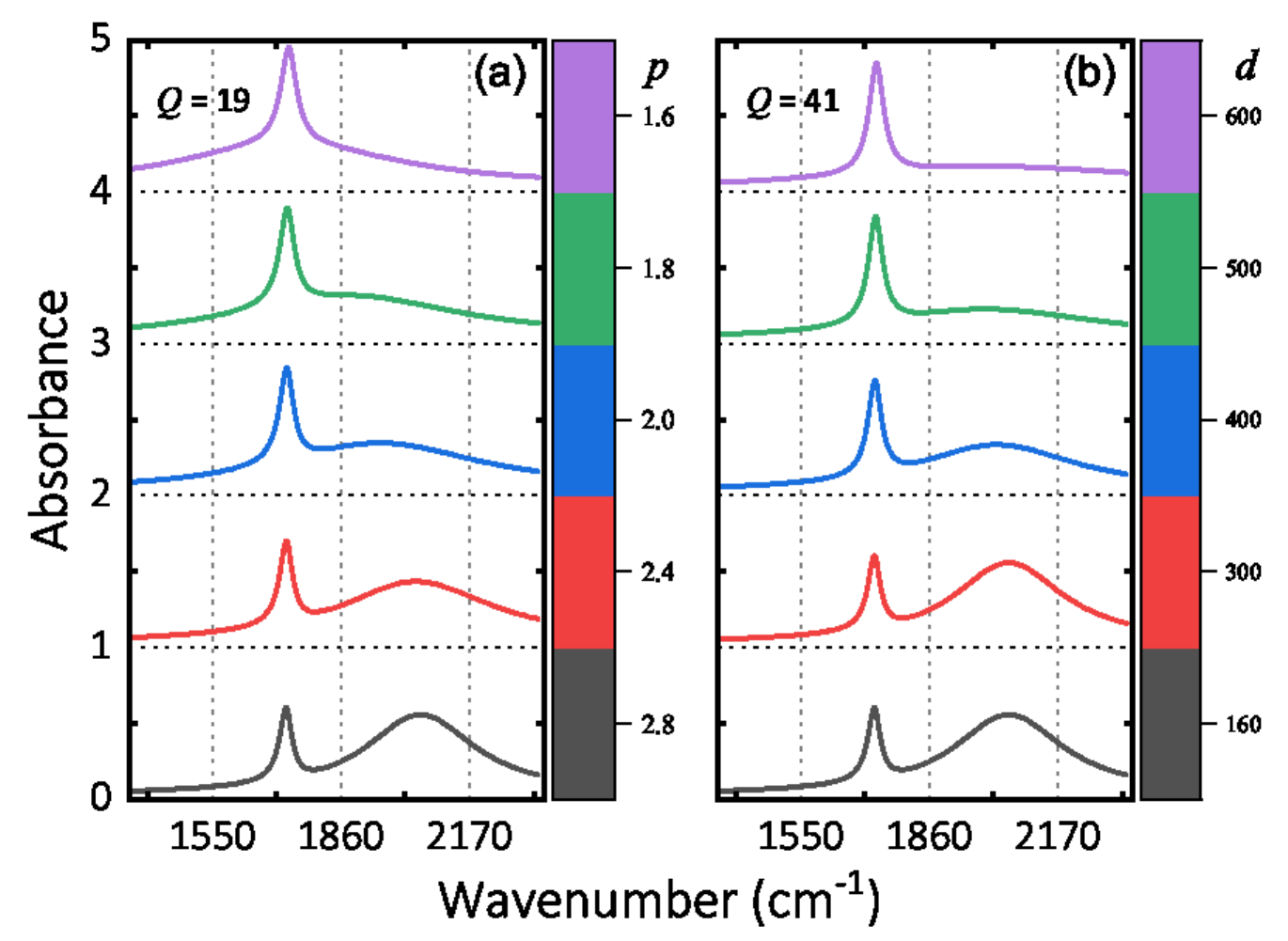}
\caption{(a) Simulated absorption spectra plotted as a function of wavelength of molecular metamaterial with PMMA film thickness of 160 nm and variable lattice constant. The arrays lattice constant varies from 2.8 to 1.6 ${\mu}$m (bottom to top). (b) Absorption spectra with a fixed period of $p$ = 2.8 ${\mu}$m and variable PMMA film thickness. The patch width was kept constant at $w$ = 1.14 ${\mu}$m for both the cases. The spectra are offset vertically for clarity.}
\end{figure}

\section{Results and discussion}
In Figure 3(a), we present the absorption spectra of the simulated molecular metamaterial for different periodicities. As shown in Figure 3(a), the absorption due to the molecular resonance increases with decreasing periodicity and reaches a maximum value of more than 98$\%$, while the absorption due to the resonance of the metamaterial decreases with decreasing periodicity. The decrease in metamaterial absorption is due to increased radiative losses, as evident from the line-width broadening of the resonance shown in Figure 3(a). We also note that the metamaterial resonance red shifts with decreasing periodicity and matches the molecular resonance for the 1.6 $\mu$m period of the array. When the two modes are spectrally overlapped, we applied the coupled-mode theory to check the critical coupling condition. The total loss rate of the system was calculated by fitting the absorption spectra with a Lorentzian fit, whereas the radiative loss rate was calculated from the simulation by treating the metal as a perfect electric conductor. From the calculation, we note that the total non-radiative loss of the system (${\gamma_{\mathrm{abs}}^{metal}}+{\gamma_{\mathrm{abs}}^{PMMA}}=44.1$)cm$^{-1}$ equals the radiative loss (${\gamma_{\mathrm{rad}}}=46$ cm$^{-1}$) of the system, which leads to the critical coupling condition \cite{Goran2015}. In this case, the reflection goes to zero, and near-perfect absorption of radiation at the molecular resonance frequency is achieved.

Next, we studied the effect of the thickness of the spacer layer, which can also be tuned to control the radiative loss of the metamaterial. As expected from Equation~5, the radiative loss of the metamaterial increases with increasing thickness of the spacer layer, leading to a broad metamaterial resonance with reduced absorption. This trend can be seen in the simulated spectra shown in Fig. 3(b). We also note that an increase in the thickness of the spacer layer red-shifts the resonance frequency of the metamaterial, which matches the C=O stretch of PMMA for 600 nm thickness. When the two resonances are tuned, the radiative environment of the metamaterial can be controlled to increase molecular absorption. We find that molecular absorption increases to a maximum of 92\% for a spacer layer thickness of 600 nm.  

\section{Conclusions}
In summary, we demonstrated high-$Q$ perfect vibrational absorption in a molecular metamaterial structure. Using a metal-insulator-metal architecture with an infrared-active polymer layer, we achieved a critical coupling condition at molecular resonance by controlling key parameters such as molecular film thickness and the periodicity of the metamaterial structure. In principle, the critical coupling condition can be achieved for any film thickness of spacer layer. As molecular absorption is an inherently narrow band compared to metamaterial resonance, perfect vibration absorption of the C = O stretch of PMMA molecules gives rise to ultra-narrow band perfect absorption with a $Q$ value of 41. Such high $Q$ factors have not been realized with conventional metal-insulator-metal architectures because of the high radiative and non-radiative losses. The perfect absorption of molecular vibrations can be of great importance in molecular sensing, spectrally selective thermal emission, single-photon emission, and filtering applications.  
 
\section*{Acknowledgement}
G. D. gratefully acknowledges the support of the Grant-in-Aid for JSPS Fellows (P20067) during his stay at the University of Tokyo. D. P. thanks IIT Delhi for the NFSG fund.

\section*{Disclosures}
The authors declare no conflict of interest.



\bibliography{references}   
\bibliographystyle{ieeetr}

\end{document}